Е.И. Хохряков

**Исследование устойчивости каскадного однобитового сигма-дельта модулятора**

*В данной статье найдена граница области устойчивости сигма дельта модулятора. Граница устойчивости зависит от внутренних коэффициентов. Граница устойчивости служит критерием для задач проектирования и сравнения различных конфигураций между собой. Определяется граница устойчивости для любого порядка, но практически найти ее можно до 5-ого порядка.*
Ключевые слова: СДМ, каскадный, устойчивость, моделирование, нелинейная модель.

E.I. Khokhryakov
**Stable equilibrium study cascaded one bit sigma-delta modulator**

In the paper defines a boundary of stability zone for sigma-delta modulator. The boundary depends from inner sigma-delta modulator coefficients. For designing purposes such result could use to find or compare some appropriate schemes with each other. It's proved some statements and showed that boundary could be found theoretically for any order of sigma-delta modulator, but practically till 5-th order.

**Введение.** Во многих приложениях применяются схемы типа сигма-дельта модулятора (СДМ). Основные направления разработки СДМ это аналогово-цифровое преобразование (АЦП) и цифро-аналоговое преобразование (ЦАП) устройства. В основу принципа функционирования СДМ заложена передискретизация [1,2], количественно характеризующаяся коэффициентом передискретизации (oversampling ratio OSR) и позволяющая изменить равномерный спектр шума квантователя на бесконечно спадающий шум при нулевой частоте. При проектировании СДМ первичной проблемой является выбор архитектуры и определение коэффициентов СДМ. В зависимости от типа устройства данная проблема решается большим числом автором по-разному. Для АЦП может быть построена поведенческая модель, позволяющая получать временные выборки и определять отношение сигнал шум в зависимости от значения OSR для СДМ тех или иных порядков. По результатам анализа поведенческой модели, как показано в работе [3], могут быть определены передаточные характеристики шума на выход из различных точек схемы. Такая методика позволяет оценить пригодность той или иной архитектуры. Другой проблемой проектирования, связанной с принципиальной реализуемостью СДМ с выбранной архитектурой и коэффициентами, является проблема анализа и обеспечения устойчивости. Как показывает поведенческое моделирование на большом числе выборок ($10^6$ и более) и при неограниченной выходной шкале интеграторов, при некоторых амплитудах входных сигналов СДМ теряет устойчивость. Изучая множество амплитуд входных сигналов, при которых теряется устойчивость, можно обнаружить несвязность этого множества, т.е. существуют окна неустойчивости внутри области устойчивости. Подобные окна неприемлемы, если речь идет об устройстве, которое должно работать в специфицированном диапазоне входной амплитуды. Окна неустойчивости были найдены для схемы из работы [4], что доказывают актуальность данной проблемы. В реальных схемах из-за конечности шкалы интеграторов для ЦАП и АЦП происходит возврат к устойчивому состоянию. Однако проблема устойчивости все же достойна внимания как при проектировании АЦП, так и ЦАП. В данной работе будет рассматриваться только случай ЦАП СДМ, встроенного в систему фазовой автоподстройки частоты (ФАПЧ) для режима дробного деления [5], при этом внутренние коэффициенты приведены к целочисленному виду, а реализация предполагает использование цифровой логики. При выборе архитектуры основные прототипы можно найти в [4] из всех возможностей главной будет однобитовый СДМ. В однобитовом СДМ нет эффекта нелинейности для ФАПЧ в дробном режиме работы. Это дает ему преимущество перед другими архитектурами, такими как многоканальный шумовой формирователь (MASH multi stage noise shaping) [6].

**Область устойчивости**. Проводя анализ области устойчивости в z-области, нужно начать с линейной модели. Решая систему уравнений относительно входного сигнала и шума квантователя, можно независимо от порядка СДМ прийти к отношению двух полиномов [2]. Коэффициенты этих полиномов будут зависеть от внутренних коэффициентов **g**, см. рис 1. Однако конкретный вид этой зависимости в дальнейшем не имеет значения

$$\text{STF}(z) = A(z)/B(z) \tag{1}$$

$$\text{NTF}(z) = C(z)/B(z) \tag{2}$$

Вид C(z) определяется спектром шума как 20n дб/дек, где n - порядок СДМ. C(z) имеет вид

$$C(z) = (z-1)^n \tag{3}$$

Про B(z) можно сказать только, что это полином n-ого порядка, старший коэффициент всегда единица. A(z) - полином первого порядка.

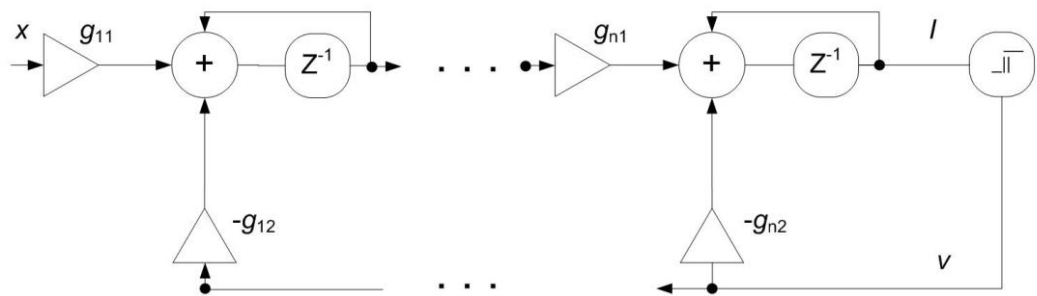

Рис 1. Структурная схема однобитового каскадного СДМ n-ого порядка.

Задача устойчивости сводится к определению диапазона выходного интегратора. Запишем основные уравнения СДМ

$$\begin{cases} v = \text{NTF}(z)E + \text{STF}(z)x \\ v = \text{sign}(I) \\ E \equiv v - I \end{cases} \tag{4}$$

где $v$ - выход компаратора, $I$ - выход последнего интегратора, $E$ - шум компаратора, $x$ - вход СДМ взяты из рис.1. Подставляя (1), (2), в (4) учитывая, что $|I| = I \cdot \text{sign}(I)$, тогда $I$ выражается как

$$I = \frac{A(z)x|I|}{|I|C(z) + \{B(z) - C(z)\}} \tag{5}$$

Заметим, что при переходе к линейному случаю нам нужно взять $|I|=1$ и E=0. В теории цифровых фильтров доказывается, что устойчивое поведение возможно только тогда, когда все полюса (5) лежат внутри единичной окружности [7]. Заметим также, что по общим свойствам полиномов B и C полином $B(z) - C(z)$ будет n-1 порядка.

Значение |*I*| в правой части (5) меняются во времени, т.е. в *z*-области это некоторая функция. Сделаем допущение, что |*I*| не меняется во времени, тогда уравнение (5) задает область устойчивости для |*I*|. Для дальнейшего будет важно определить границы этой области как функцию коэффициентов полинома B(*z*), для этой цели такое допущение оправдано.

**Необходимый и достаточный критерий для подсчета нулей.** В теории функции комплексной переменой (ТФКП) доказывается следующие два утверждения. Первое, изменение аргумента аналитической функции F(*z*) вдоль контура определяется числом $2i\pi n$, где n-число нулей внутри контура. Для доказательства достаточно проинтегрировать по указанному контуру функцию F(*z*)'/F(*z*). Второе, приводится теорема Руше, которая совместно с неравенством треугольника утверждает достаточные условия при подсчете количества нулей полинома, находящиеся внутри некоторого контура [9].

На рис.2 рассмотрен полином, который этому условию не удовлетворяет и все же его корни находятся внутри контура единичной окружности, поэтому применение теоремы Руше не необходимо, а, значит, возникает потребность сформулировать необходимое и достаточное условие.

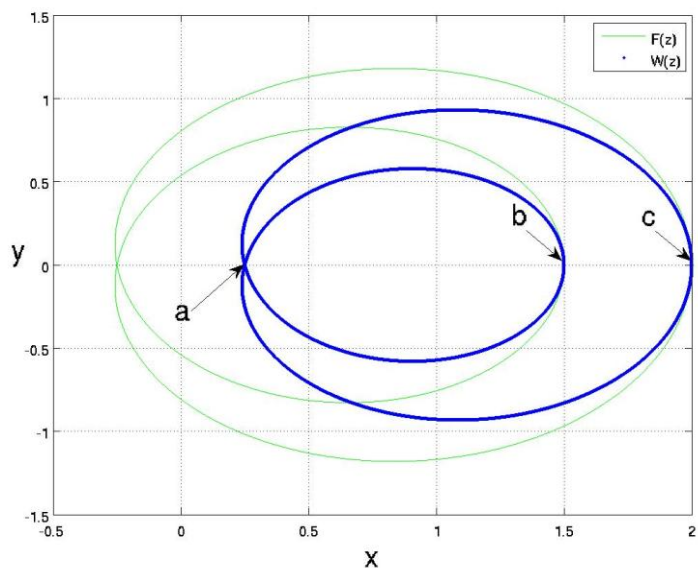

Рис 2. Контур $|z|=1$ для полинома W($z$) - точки и F($z$) = $z^2 + 1/2z + 3/4$ - линии.

Из рис.2 видно, что для контура F($z$), где $|z|=1$ имеется 2 петли вокруг нуля, что подтверждает вышеуказанное утверждение. Построим также функцию W($z$) = F($z$)/f($z$). Ясно, что аргумент W($z$) будет равен разности аргументов F($z$) и f($z$), тогда количество нулей F($z$) и f($z$) совпадает только в том случае, если у W($z$) нет нулей. Применяя первое утверждение ТФКП, справедливо следующее:

*E: Количество нулей F(z) и f(z) совпадает тогда и только тогда, когда отображение контура $|z|=1$ функцией W(z) не содержит нулей.*

Функция W($z$) есть полином $z^{-1}$, однако ничто не мешает его рассматривать как полином от $z$, поскольку изменение знака степени лишь меняет направление обхода контура.

На рис 2 кривая W($z$), для $|z|=1$ как раз находится вне нуля. Чтобы преобразовать утверждение E для численных расчетов достаточно заметить, что на рис. 2 есть характерные точки при $\text{Im}(F(z))=0$. Эти точки показывают или поворот контура для $z=\pm1$ (эти точки называются в дальнейшем перманентными см рис 2, точки *b* и *c*), или точки самопересечения контура (так будут называться эти точки в дальнейшем см рис 2, точка *a*). Характеристические точки лежат на действительной оси, поскольку все коэффициенты полинома W($z$) действительны.

Сформулируем два вспомогательных утверждения

*L1: Для W(z) точки самопересечения и перманентные точки лежат на действительной оси.*

Для доказательства L1 нужно показать, что перманентные точки всегда существуют. Это понятно, поскольку они являются решением $z=\pm1$, т.е. аргумент, $\phi=0,\pi$, который всегда будет корнем $\text{Im}(W(z))=0$ для действительных коэффициентов.

Теперь нужно показать, что если точки самопересечения существуют, то они могут находиться только на действительной оси. Это следует из того, что нужно представить точки пересечения как корни $R(x) = \text{Im}(W(z))/\sqrt{1-x^2} = 0, |x| < 1$ где $x = \text{Re}(z)$ - действительное число. Подставляя корни R(x) для Re(W(z)), получим действительное значение. Отметим также, что если корней R(x) не существует, то и точек самопересечений контура W(z) также не будет.

*L2: Точки самопересечения, если они существуют, не могут находиться между перманентными точками.*

Для доказательства L2 заметим, что все характерные точки проходятся при изменении аргумента от 0 до $\pi$. Далее от $\pi$ до $2\pi$ проходим по симметричной части контура от противоположной перманентной точки, замыкая все внутренние петли через точки самопересечения. Более детализированное построение различных контуров показывает, что в перманентных точках контур поворачивает всегда только в одну сторону, а петель в виде восьмерки не наблюдается.

Для того, чтобы не было петлей вокруг нуля необходимо, чтобы перманентные точки лежали всегда справа от 0. В силу утверждения L2 точки самопересечения могут лежать левее перманентных, поэтому для того, чтобы не было петель вокруг 0, а значит нулей W(z) нужно, чтобы они были все же больше нуля. Однако W(z) может не иметь нулей в случае, когда обход внешней петли противоположен внутренней. Теперь есть все, чтобы сформулировать следующее утверждение

*E1: Полином n-ого порядка F(z) будет иметь n-нулей внутри контура $|z|=1$ тогда и только тогда, когда для всех z, определяемые из $\text{Im}(W(z)) = 0$, справедливо $\text{Re}(W(z)) \leq 0$ только для четного количества точек самопересечения, а перманентные точки лежат справа от нуля.*

Для доказательства важно понимать, что направление обхода первой внутренней петли совпадает с внешней. Далее направление следующих петель чередуется по мере приближения к левой перманентной точке. Тогда петля, охватывающая 0, будет четной и, значит, будет направлена противоположно внешней.

Следствие 1. Может показаться, что условия E1 не выполнены, когда нет ни одной петли вокруг нуля. Все точки самопересечения лежат правее нуля или их не существует вообще, даже для положительного Re(W(z)). Для этого случая достаточно вспомнить, что нулевое количество точек самопересечения это также четное число.

Следствие 2. В E1 требуется, чтобы перманентные точки всегда были положительны. Однако также подходящим случаем будет, если перманентные точки окажутся отрицательными и при этом $Re(W(z)) \geq 0$. Без ограничения общности считаем, что перманентные точки располагаются всегда только справа от 0, поскольку для W(z) знак минус всегда можно вынести.

Следствие 3. Коэффициенты полинома W(z) для СДМ зависят от |I|, поэтому момент, когда точка самопересечения проходит через начало координат, очень важен. Вне зависимости влево или вправо движется точка самопересечения при изменении |I|, в момент прохождения нуля меняется характер устойчивости СДМ, т.е. устойчивость переходит в неустойчивость и наоборот.

**Условия для нулевой точки.** Нулевой точкой будем называть точку самопересечения контура W(z), $|z|=1$, находящуюся в нуле. Следствия 3 предыдущего пункта позволяет судить о нулевой точке как точке смены устойчивости. В этом пункте найдется условие, для |I| при, котором реализуются нулевые точки самопересечения.

По определению нулевой точки необходимо приравнять к нулю мнимую и действительную часть, при этом для мнимой части нужно выбросить перманентные корни. Вводятся полиномы с аргументом на интервале (-1,1) R0 и R1

$$\begin{cases} R_0^n(x) \equiv Re(W(z)) = 0 \\ R_1^{n-1} \equiv Im(W(z))/\sqrt{1-x^2} = 0, |x|<1, x = Re(z) \end{cases}$$

Здесь верхние индексы R указывают порядок полинома. Нужно представить R0 в виде целой и дробной части от R1. Ясно, что поскольку целая часть R0 обращается в ноль, то дробная тоже. Обозначая эту часть как R2, заметим, что для R2 порядок уменьшится на 1 по сравнению с R1. Можно продолжить брать дробную часть предпоследнего от последнего остатка до тех пор, пока не получится полином нулевого порядка. Приравняв этот последний остаток к нулю, можно выразить |*I*| от коэффициентов **b**. Для наглядности верхний и нижний индекс перейдет в первый и второй аргумент соответственно, т.е.

$R_k^{n-k}(x) \equiv R[n-k, k]$

$$R[n-k, k] = \mathrm{mod}(R[n-k+2, k-2], R[n-k+1, k-1]) = 0, k = 2, 3, \ldots, n \qquad (6)$$

Выражение (6) для $k = n$ приводит к тому же результату, если корни R1 подставить в R0. Однако представляют интерес не все корни, а только те, которые вещественны и меньше по модулю 1, также не интересует случай, когда старший коэффициент R0 обращается в ноль. Последний случай соответствует исчезновению корня, поскольку порядок уменьшается.

**Определение границ устойчивости.** Для определения области устойчивости |*I*| пользуемся E1 для полинома знаменателя (11). Для этого берется из E1 условие для перманентных точек. Для другой части E1 вместо того, чтобы искать точки самопересечения и считать их число левее нуля, можно воспользоваться следствием 3 и (6). Коэффициенты **b** в дальнейшем сопоставляются полиному B(z)-C(z). Тогда

$$\begin{cases} |I| + \sum_{k=1}^{n} \{b_k + C_n^k (-1)^k |I| \cos(k\phi)\} > 0, \phi = 0, \phi = \pi \\ \sum_{k=1}^{n} \{b_k + C_n^k (-1)^k |I| \sin(k\phi)\} = 0, \sin(\phi) \neq 0 \\ |I| + \sum_{k=1}^{n} \{b_k + C_n^k (-1)^k |I| \cos(k\phi)\} = 0, \sin(\phi) \neq 0 \end{cases} \qquad (7)$$

Где $C_n^k$ - бином Ньютона. *k*-ый аргумент синуса и косинуса раскрывается по рекуррентной формуле

$$\begin{cases} \sin(k\phi) + \sin((k-2)\phi) = 2\sin((k-1)\phi)\cos(\phi) \\ \cos(k\phi) + \cos((k-2)\phi) = 2\cos((k-1)\phi)\cos(\phi) \\ k > 2 \end{cases} \qquad (8)$$

В результате этого преобразования для 2, 3 уравнения (7), соответственно, все сводится к наименьшим аргументам $\phi$ и $2\phi$. В таком представлении соберется в качестве множителей два полинома $T_1(\cos(\phi))$ и $T_2^{n-2}(\cos(\phi))$ перед одинарным и удвоенным аргументом соответственно. Индекс сверху полинома Т2 означает его порядок для положительных значений, а для отрицательных значений порядок берется нулевой. В качестве аргументов Т1 и Т2 получились косинусы для обоих уравнений. Вводя переменную $x = \cos(\phi)$, (7) запишется как

$$\begin{cases} \sum_{k=1}^{n} b_k > 0 \\ |I|\min = -\sum_{k=1}^{n}(-1)^k b_k / 2^n \\ 2x\,T_2(x) + T_1(x) = 0, |x| < 1 \\ |I| + \{2x\,T_2(x) + T_1(x)\}x = |I| - T_2(x) = 0, |x| < 1 \end{cases} \tag{9}$$

Система (7) определяет границы области устойчивости. Можно видеть, что нижняя граница уже получена.

**Полином Т2.** Для решения 4-ого уравнения (9) выпишем полиномы до 5-ого порядка. Делается замена $d_k = b_k + C_n^k(-1)^k|I|$. Удобно получить полиномы для синуса и косинуса в среде Matlab как

syms d1 d2 d3 d4 d5 x

ps=d5*sin(5*acos(x))+d4*sin(4*acos(x))+d3*sin(3*acos(x)) …

+d2*sin(2*acos(x))+d1*sin(acos(x))

pc=d5*cos(5*acos(x))+d4*cos(4*acos(x))+d3*cos(3*acos(x)) …

+d2*cos(2*acos(x))+d1*cos(acos(x))

pC=simplify(ps), pS=simplify(ps)

далее из полученных полиномов нужно взять остаток при делении косинус полинома pC на синус полином pS, тогда

$$T_2(x) = -(-8d_5 x^3 - 4d_4 x^2 + \{4d_5 - 2d_3\}x - d_2 + d_4) \tag{10}$$

**Определение верхней границы устойчивой области.** Верхняя граница получается из (6) при подстановке двух последних уравнений (9). Третье уравнение (9) получается, см. выше, попутно вместе с (10). Применяя (6), возникает проблема - лишние корни, которые получаются при подстановке не правильных корней для косинус полинома в синус полином. К числу неправильных корней относятся все комплексные корни или корни по модулю большие 1. Однако трудности с лишними корнями возникают только для пятого порядка СДМ и выше.

Для конкретного случая можно получить результат для 3-его порядка

$$|I|\max = (b_1 b_3 - b_3^2)/(b_1 + b_2 + b_3) \qquad (11)$$

**Выводы.** В статье рассмотрена область устойчивость для каскадного однобитового СДМ любого порядка. Рассматривая СДМ как ЦАП прототип, реализованный в цифровой логике, фактически получается одна реализация. В отличие от АЦП [9] реализация ЦАП может быть точно промоделирована и не требует отдельных измерений, поскольку внутренние коэффициенты не подвержены искажениям. Для численных оценок при вариации внутренних коэффициентов моделирование дает только шумовые характеристики, задающие эффективную разрядность и OSR. Для исследования устойчивости в работе представлена принципиально другая характеристика – граница устойчивости. Получены соотношения (9) на основе которых можно проанализировать верхнюю и нижнюю границу устойчивости СДМ любого порядка. В частности получена граница для СДМ 3-его порядка (11). Для ФАПЧ в дробном режиме 3-его порядка был изначально предложен вариант из [4], на выборках из $10^6$ выборок обнаружена потеря устойчивости. Незначительная модификация приведенных к целочисленному виду коэффициентов СДМ привело к полному исчезновению окон неустойчивости, таким образом, найдены лучшие коэффициенты, чем в [4] . Вместе с тем граница устойчивости также расширилась. Аналогичная методика применена и для более высоких порядков.

**Литература**

1. Schreier R., Temes G., Understanding delta-sigma data converters// NJ: IEEE Press, Piscataway, 2005, P. 1-10

2. Pervez M., Henrik V., Jan Van der Spiegel, An overview of sigma-delta converters: how a 1-bit ADC achieves more than 16-bit resolution // IEEE Signal Processing Magazine, Jan 1997, P. 61-84

3. Гусев В.В., Кондратенко С.В., Скок Д.В. Анализ влияния неидеальностей характеристик блоков при проектировании высокоразрядных сигма-дельта модуляторов// Материалы 11-й всероссийской научно-технической конференции: Электроника, микро и наноэлектроника. Сб. научн. трудов. – М.: МИФИ, – 2009. – Т. 1. – С. 167-178.

4. Mingliang L, Tutorial on Designing Delta-Sigma Modulators// Extron Electronics, 2004, www.commsdesign.com/showArticle

5. Kundert K, Predicting the Phase Noise and Jitter of PLL-Based Frequency Synthesizers// The Designer's Guide Community Forum. Available from www.designersguide.org

6. Shu K., Sunchez-Sinnencio.E. Franco M, Udaykiran. E., A comparative study of digital sigma-delta modulators for fractional-N-synthesis. // Proc. IEEE Int. Conf. Electronics, Circuits, and Systems (ICECS), 2001, P.1391 -1394

7. Скляр Б. Цифровая связь. Теоретические основы и практическое применение//Вильямс,М,изд 2, 2003, P. 1079-1080

8. Тихонов АГ, Свешников АН, Теория функций комплексного переменного// Наука.,М. с. 144-147

9. Lucien J. Breems, Robert Rutten, Gunnar Wetzker, A cascaded continuous-time ΣΔ modulator with 67-dB dynamic range in 10-MHz bandwidth// IEEE JOURNAL OF SOLID-STATE CIRCUITS, VOL. 39, NO. 12, DECEMBER 2004, P. 157